\documentclass[twocolumn,showpacs,showkeys,amsmath,amssymb]{revtex4}

\usepackage{graphicx}
\usepackage{dcolumn}
\usepackage{bm}
\begin{document}


\title{Fast nucleon emission as a probe of the isospin 
momentum dependence}

\author{J. Rizzo, M. Colonna, M. Di Toro}
\affiliation{Laboratori Nazionali del Sud INFN, Via S.Sofia 62, I-95123
Catania, Italy\\ 
Dipartimento di Fisica e Astronomia, Universita' di Catania\\
E-mail: colonna@lns.infn.it}

\begin{abstract}

In this article we investigate the structure of the non-local
part of the symmetry term, that leads to a splitting of the effective
masses of protons and neutrons in asymmetric matter.   
Based on microscopic transport simulations we suggest
some rather sensitive observables in collisions of neutron-rich (unstable)
ions at intermediate ($RIA$) energies.
In particular we focus the attention on pre-equilibrium nucleon emissions.
We discuss 
interesting correlations between the $N/Z$ content of the fast emitted 
particles
and their rapidity or transverse momentum, that show a nice dependence
on the prescription used for the effective mass splitting.

\end{abstract}

\maketitle
\section{Introduction}

Effective interactions have been extensively used in transport codes to gain
knowledge about nuclear matter properties in conditions far from equilibrium.
They predict very different behaviours for some physical properties, such as 
the
momentum dependence of the mean field in the iso-scalar and iso-vector channel
(see \cite{Isospin01,BaranPR410} for recent reviews).

In general, the momentum dependence of the potential is strictly related to
the concept of in-medium reduction of the nucleon mass. 
In particular, the non-local part of the symmetry potential has become 
an interesting subject of investigation, because it leads to a splitting of the
masses of different nucleonic species in asymmetric matter, 
and the sign of this splitting already is
an open and controversial problem (\cite{RizzoNPA732,BaoPRC69,DiToroarXiv}).

Considering separately the local and non-local contributions to the
potential part of the symmetry energy, 
a sharp change is observed, for instance, going 
from the earlier Skyrme forces to the Lyon parametrizations, with 
almost an inversion of the signs of the two contributions 
 \cite{BaranPR410}. The important
repulsive non-local part of the Lyon forces leads to a completely
different behavior of the neutron matter $EOS$, of great relevance 
for the neutron star properties. Actually this substantially modified
parametrization was mainly motivated by a very unpleasant feature
in the spin channel of the earlier Skyrme forces, the collapse of
polarized neutron matter
\cite{KutscheraPLB325,LyonNPA627,LyonNPA635,LyonNPA665}.
In correspondence the
predictions on isospin effects on the momentum
dependence of the symmetry term are quite different.
A very important consequence for the reaction dynamics is the
expected inversion of the sign of the $n/p$ effective mass splitting, i.e. 
in the Lyon forces neutron effective
masses are below the proton ones for n-rich matter.

We note that the same is predicted
from microscopic relativistic Dirac-Brueckner-Hartree-Fock ($DBHF$) 
calculations
\cite{HofmannPRC64,SchillerEPJA11,DalenNPA744} and in general from the 
introduction of 
scalar isovector
virtual mesons in Relativistic Mean Field
 ($RMF$) approaches \cite{LiuboPRC65,GrecoPRC67}. At variance, 
non-relativistic Brueckner-Hartree-Fock ($BHF$)
calculations are leading to opposite conclusions
 \cite{Bombiso,ZuoPRC60,ZuoPRC72}. However, the comparison between 
relativistic
effective ($Dirac$) masses and non-relativistic effective masses requires
some attention, see the Ch.6 of \cite{BaranPR410} and refs. therein. 
In fact, though the neutron $Dirac$-mass is below the proton one,  
in correspondence the non-relativistic $Schroedinger$-mass 
splitting can have any sign. Indeed this is just the case of the very
recent $DBHF$ analysis of the T\"ubingen group \cite{DalenNPA744,Dalenar0502}.
Such puzzling effect is related to the intrinsic momentum dependence
of the nucleon self-energies, it is then beyond the $RMF$ approach and
it represents a very sensitive test of the treatment of Dirac-Brueckner
correlations. All that clearly shows that sensitive experimental observables
are largely needed.

The sign of the splitting will directly affect the energy
dependence of the so-called Lane Potential, i.e. the difference between 
$(n,p)$ optical
potentials on charge asymmetric targets, normalized by the target asymmetry
\cite{LaneNP35}. A decreasing behaviour is observed in the case $m^*_n>m^*_p$, 
while a positive slope is obtained in the opposite case. 
An important physical consequence of the negative slopes is that the isospin
effects on the optical potentials tend to disappear at energies just above
$100~MeV$ (or even change the sign for ``old'' Skyrme-like forces).
Also, we expect a crossing of the two prescriptions   
at low energies, i.e. low momentum
nucleons will see exactly the same Lane potentials, as shown in detail
in ref. \cite{DiToroarXiv}.

Unfortunately results derived from neutron/proton optical potentials at 
low energies
are not conclusive \cite{LaneNP35,BecchettiPR182,Hodopt}, since the 
effects due to the mass splitting appear
of the same order of the uncertainty on the determination of the local
contribution to the symmetry energy. Moreover at low energies, due to the 
crossing discussed before, it is difficult to appreciate 
differences between positive or negative slopes. 

A positive slope is obtained, for instance, within
the phenomenological
Dirac Optical Potential ($Madland-potential$), 
with different implicit momemtum
dependences in the self-energies  \cite{KozackPRC39,KozackNPA509}. 
This potential has been constructed fitting simultaneously proton and
neutron (mostly total cross sections) data for collisions with a wide
range of nuclei at energies up to $100~MeV$. Recently such Dirac optical
potential has been proven to reproduce very well the new neutron 
scattering data on $^{208}Pb$ at $96~MeV$ \cite{KlugPRC6768} measured at
the Svendberg Laboratory in Uppsala.

More data are needed at higher energies (around/above $100~MeV$), 
to improve the  
systematics, in order to clearly disentangle between the two trends
of the Lane potential and of the effective mass splitting.  

We note however that these properties of the interaction 
will also affect the dynamical 
evolution of
heavy ion collisions. So we can get indipendent information about it 
just by looking at some
suitable reaction observables.
We can expect important effects
on transport properties ( fast particle emission, collective flows)
 of the dense and asymmetric $NM$ that will
be reached in Radioactive Beam collisions at intermediate energies.

Here we will focus on the study of pre-equilibrium emission in neutron-rich 
central
collisions, at 50 and 100 MeV/A. We will see that the energy dependence of 
the N/Z content 
of fast emitted particles is particularly sensitive to the sign of the 
effective mass
splitting. Hence it 
would be possible to answer many fundamental questions in isospin physics
by looking at appropriate observables in intermediate energy $HIC$.

\section{Details of the interaction}

In order to directly test the influence of the $Schroedinger$-effective mass
splitting we will present results from reaction dynamics at intermediate
energies analysed in a non-relativistic transport approach, of $BNV$
type, see \cite{Alfio,BaranPR410}. The $Iso-MD$ effective interaction is
derived via an asymmetric extension of the $GBD$ force
 \cite{GalePRC41,GrecoPRC59}.

The energy density  can be parametrized
as follows (see also \cite{Isospin01},\cite{Bombiso}):
\begin{equation}
\varepsilon=\varepsilon_{kin}+\varepsilon(A',A'')
+\varepsilon(B',B'')+\varepsilon(C',C'')
\label{edensmd1}
\end{equation}
where $\varepsilon_{kin}$ is the usual kinetic energy density and
\begin{eqnarray}
\varepsilon(A',A'')=(A'+A''I^2)\frac{\varrho^2}{\varrho_0}
 \nonumber \\
\varepsilon(B',B'')=(B'+B''I^2)\left ( \frac{\varrho}{\varrho_0}
\right )^{\sigma}\varrho
 \nonumber \\
\varepsilon(C',C'')=C'(\mathcal{I}_{NN}+\mathcal{I}_{PP})
+C''\mathcal{I}_{NP}
\label{edensmd2}
\end{eqnarray}
The variable $I=(N-Z)/(A)$ defines the isospin content of the system,
given the number of neutrons $(N)$, protons $(Z)$, and the total mass $A=N+Z$;
the quantity $\varrho_0$ is the normal density of nuclear matter.
The momentum dependence is contained in the $\mathcal{I}_{\tau \tau'}$
terms, which indicate integrals of the form:

$$\mathcal{I}_{\tau \tau'}=\int d \vec{p} \; d \vec{p}\,' 
f_{\tau}(\vec{r},\vec{p})
 f_{\tau'}(\vec{r},\vec{p}\,') g(\vec{p},\vec{p}\,')$$
with $g(\vec{p},\vec{p}\,')=(\vec{p}-\vec{p}\,')^2$.
This choice of the function $g(\vec{p},\vec{p}\,')$ corresponds to
a Skyrme-like behaviour 
and it is suitable for $BNV$ simulations. 
We use a soft equation of state for symmetric nuclear matter 
(compressibility modulus $K_{NM}(\varrho_0)=215 \,MeV$).
In this frame
we can easily adjust the parameters in order to have the same density
dependence of the symmetry energy 
{\it but with two opposite n/p effective mass splittings}, as predicted by the
early Skyrme forces \cite{LyonNPA627} and the later Skyrme-Lyon parametrizations
\cite{LyonNPA635}. So we can 
separately study the corresponding dynamical effects, \cite{RizzoNPA732}. 

\begin{figure}[hb]

\begin{picture}(0,0)
\put(110.5,25){\mbox{\includegraphics[width=3.6cm]{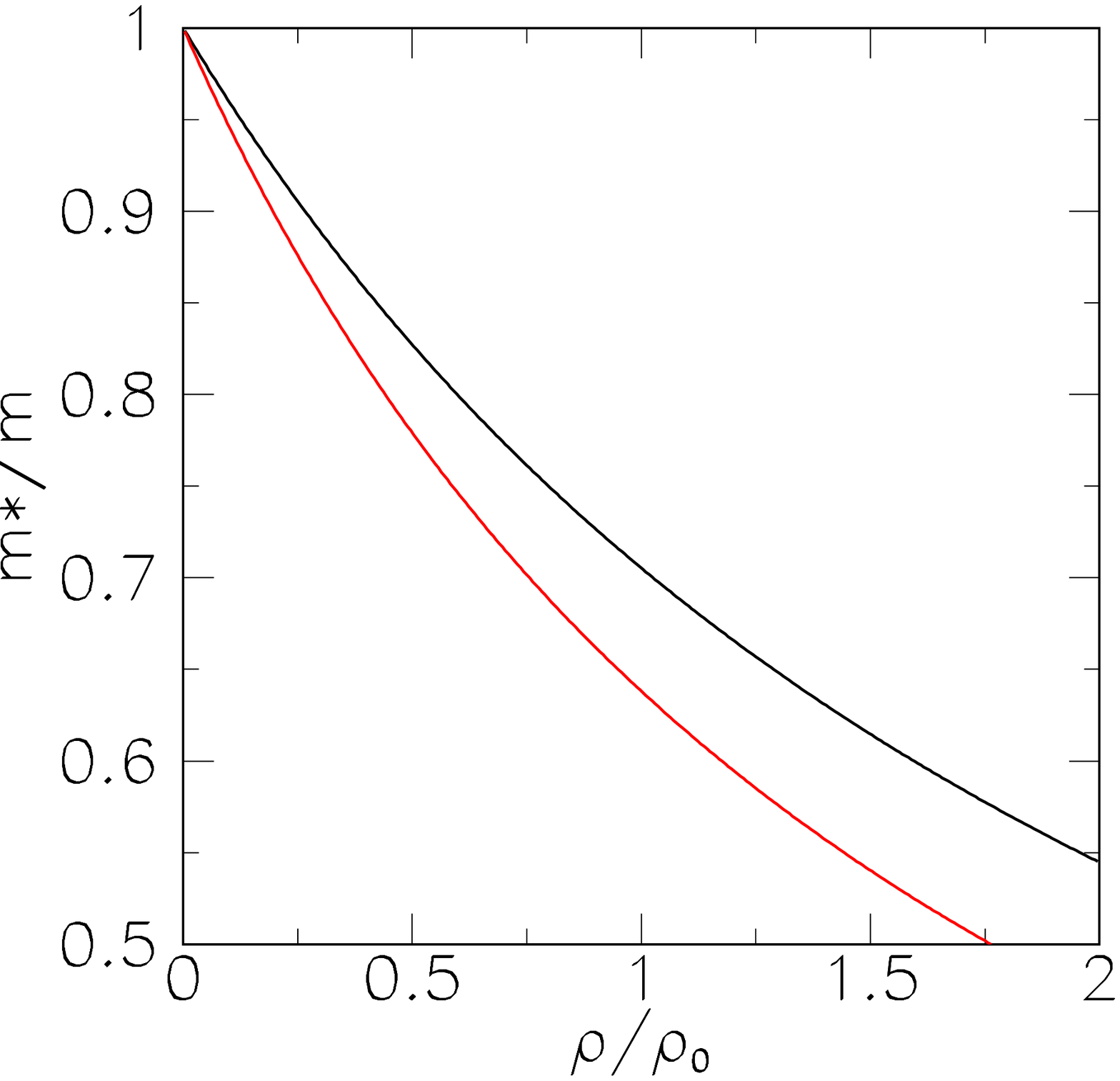}}}
\end{picture}
\includegraphics[width=6.cm]{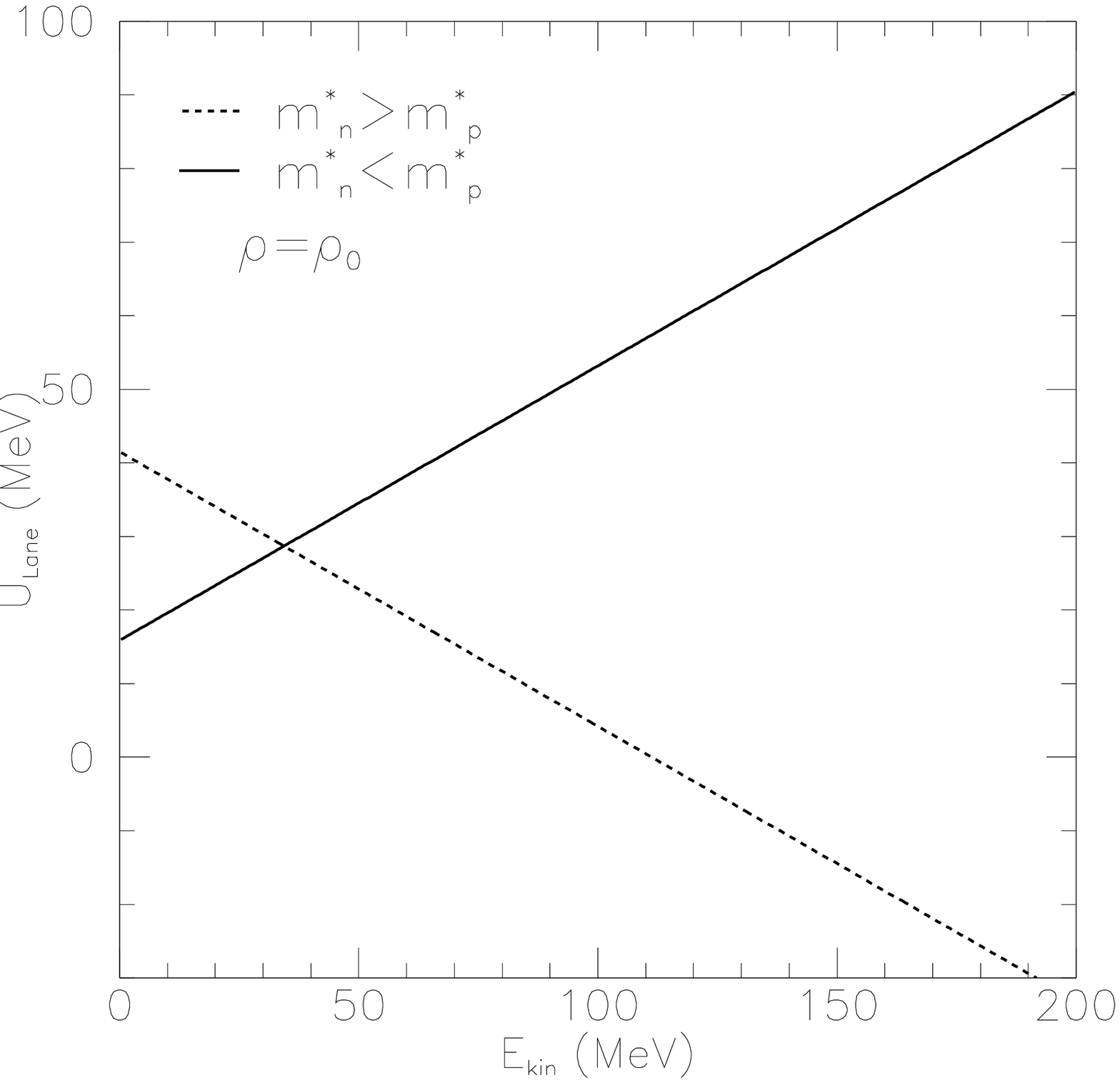}

\caption{\it Energy dependence of the Lane potential for the used 
parametrizations. 
Small panel: the effective mass splitting, related to the slope of the Lane
potential (for a $I=0.2$ asymmetry).} 
\label{lanepseudo}
\end{figure}

A good
way to visualize the physical meaning of the mass splitting is to look at the
kinetic energy dependence of the
Lane potential $U_{Lane}=(U_n-U_p)/2I$, \cite{LaneNP35}. 
Its slope depends on the value and sign of the mass splitting (see
\cite{DiToroarXiv} for details). 
In Fig. \ref{lanepseudo} we plot the Lane potential
with the parameters corresponding to the two choices of the sign
of the n/p mass splitting (shown in the insert for the $I=0.2$ asymmetry).
In the two cases, the absolute value of the splitting 
is exactly the same (the difference between neutron and proton masses is
$\sim 10 \% $ 
at normal density), only the sign is opposite.
The upper curve well reproduces the Skyrme-Lyon (in particular $SLy4,Sly7$)
results, the lower (dashed) the $SIII,SKM^*$ ones.

From the figure we can immediately derive the expectation of very different
symmetry effects for nucleons around $100MeV$ kinetic energy:
enhancement (with larger neutron repulsion) in the  $m_n^*<m_p^*$ case
vs. a  disappearing (and even larger proton repulsion) in the
$m_n^*>m_p^*$ choice.
Besides, we can see a crossing of the two prescriptions 
at low energy: this means that
low momentum nucleons feel almost the same Lane potential. Thus it is important
 to
choose the appropriate dynamical observables in order to clearly see the effect
of mass splitting, e.g. to look at observables
where neutron/proton mean fields at high momentum are playing an
important role.


\section{Isotopic content of the fast nucleon emission}

We have performed realistic ``ab initio'' simulations of collisions
at intermediate energies of n-rich systems, in particular 
 $^{132}Sn+^{124}Sn$ at $50$ and $100\, AMeV$, at impact parameter $b=2\,fm$. 
We have employed either 
an $asy-stiff$ or an $asy-soft$ density dependence of 
the symmetry energy, corresponding to different slopes around
normal density, with an increasing symmetry repulsion in the $asy-stiff$
 case, \cite{BaranNPA703}. One of the aims of our work is also to
select reaction observables more sensitive to the momentum (non-local part)
than to the density (local part) dependence of the symmetry term.
 
\begin{figure}[htb]
\includegraphics[width=8.cm,height=7.cm]{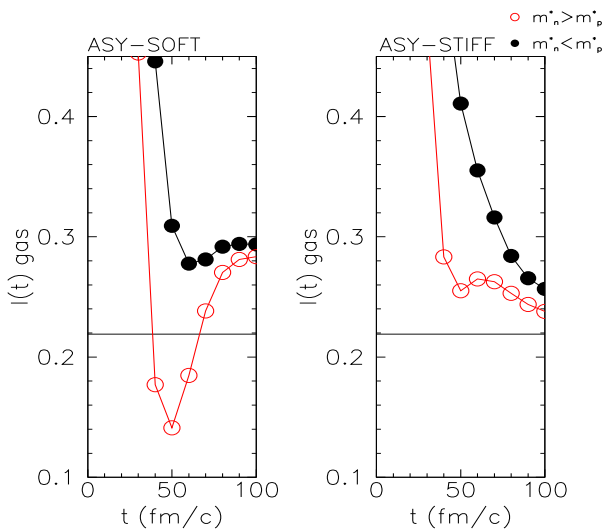}
\caption{\it Isospin gas content as a function of time in a central collision 
$^{132}Sn+^{124}Sn$ at $50 \,AMeV$ for two opposite choices of mass splitting.
Left panel: asy-soft EOS; right panel: asy-stiff EOS.}
\label{asyevol}
\end{figure}

Performing a local low density selection of the test particles
($\rho<\rho_0/8$) we can follow the time evolution of nucleon
emissions (gas phase) and the corresponding asymmetry.
It is shown in Fig. \ref{asyevol} 
 (where $I(t) \equiv (N-Z)/A$) for the $50\,AMeV$ reaction.
 The solid line gives the initial asymmetry of the total system. 
 
 A general feature of nucleon
 emission is the isospin distillation 
 (see \cite{BaranNPA703} and refs. therein), which leads to a gas
 phase with an asymmetry greater than the initial one. We find an interesting 
 behaviour for the isospin content of the nucleonic gas:
  at early times (up to about 60 fm/c), during the (high
density) compression phase, the asy-stiff choice leads 
to higher isospin asymmetry than the asy-soft one; 
later, we can see an inversion of this trend, due to large contributions to the
gas phase coming from low density regions \cite{BaranNPA703};
finally, the gas phase is more asymmetric in the asy-soft case 
\cite{BaoINTJE7}.

This behaviour is mainly due to the local part of the symmetry energy, but
for both choices of the density stiffness of the symmetry
term we clearly see also the effects of the mass splitting, resulting in
a reduced fast neutron emission when $m_n^*>m_p^*$. It is more pronounced 
during
the early stages of the reaction (up to $60\, fm/c$), when the most energetic
particles are emitted \cite{ChenPRC68} and the momentum dependent
part of the mean field is more effective. The two effects we have just 
discussed,
the former related to the density dependence, 
the latter to the momentum dependence
 of the symmetry potential, can be singled out in the
transverse momentum distribution of the N/Z content of the gas phase.

\begin{figure}[thb]
\centering
\includegraphics[scale=0.5]{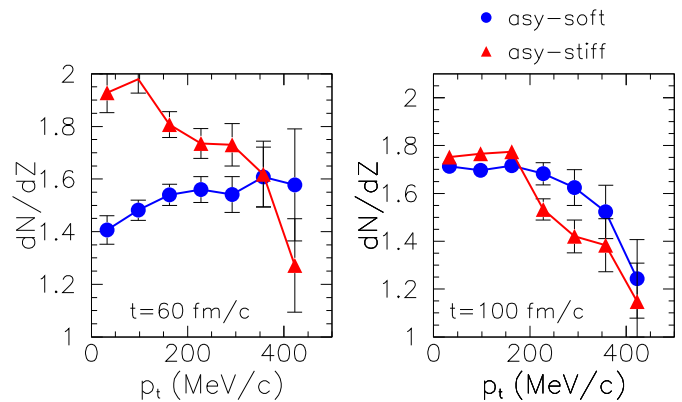}
\caption{\it Transverse momentum dependence of the neutron/proton
ratio in the rapidity range $|y^{(0)}| \leq 0.3$ for a central reaction 
$^{132}Sn+^{124}Sn$ at $50 \,AMeV$. Comparison between two choices of symmetry
energy stiffness at the two different times defined in the text.}  
\label{sn132_50}
\end{figure}

\begin{figure}[tbp]
\includegraphics[scale=0.5]{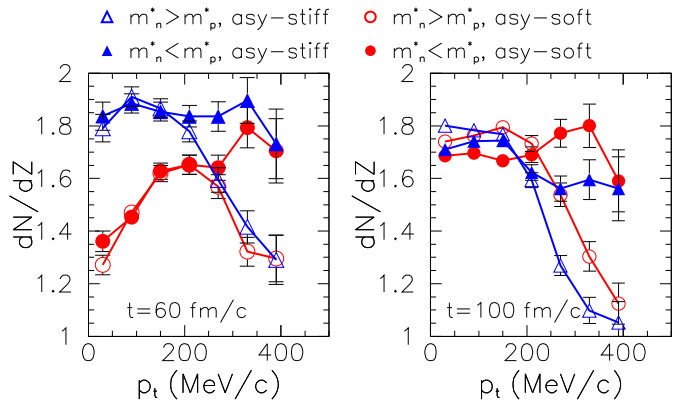}
\caption{\it Same as Fig. \ref{sn132_50}, for two opposite choices of mass
splitting.} \label{sn132_50_split}
\end{figure}

In order to better isolate the mass splitting effect and to select
the corresponding observables, in Figs. \ref{sn132_50},~\ref{sn132_50_split}
 we report the $N/Z$ of the ``gas'' at two different times, $t=60fm/c$ (end 
 of the pre-equilibrium emission), and $t=100fm/c$ (roughly freeze-out 
time). We have followed the transverse momentum dependence, for
 a  fixed central rapidity $y^{(0)}$ (normalized to projectile rapidity).\\
In Fig. \ref{sn132_50} we show the results {\it without the mass splitting
effect}, i.e. only taking into account the different repulsion of the
symmetry term (the initial average asymmetry is $N/Z=1.56$).

At early times (left panel of Fig. \ref{sn132_50}), when the emission 
from high density regions is
dominant, we see a difference due to the larger neutron repulsion
of the asy-stiff choice. The effect is reduced at higher $p_t$'s
due to the overall repulsion of the isoscalar momentum dependence.
Finally at freeze out (right panel)
the difference is less pronounced, since 
$Isospin~Distillation$ is coming into play during the expansion phase; 
it is more efficient in the asy-soft choice,
 \cite{BaranPR410} and refs. therein.

When we introduce the mass splitting, the difference in the isotopic content
of the gas, for the larger transverse momenta, is very evident at all times,
 in particular at the freeze-out of experimental interest, 
see Fig. \ref{sn132_50_split}.
As expected the $m_n^*>m_p^*$ sharply reduces the neutron emission
at high $p_t$'s. On the other hand, low momentum nucleons show a 
behaviour which
is consistent with Fig. \ref{sn132_50}, mainly ruled by the local part 
of the symmetry potential.


\begin{figure}[htbp]
\includegraphics[scale=0.5]{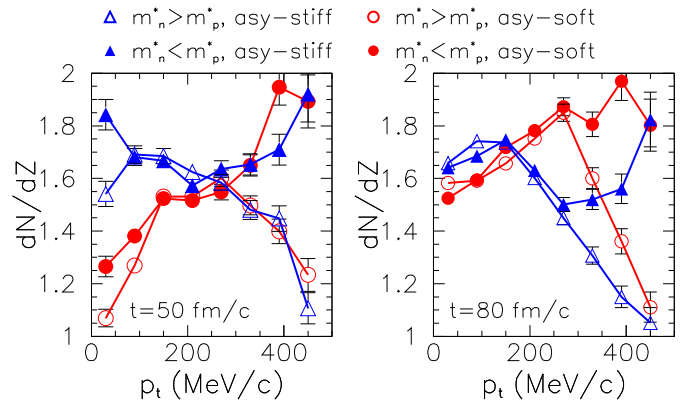}
\caption{\it Same as Fig. \ref{sn132_50_split}, but for an incident energy of 
$100\, AMeV$ (note the reduced ``pre-equilibrium'', left panel, and
 ``freeze-out'', right panel, times).} \label{sn132_100_split}
\end{figure}

We have repeated the analysis at higher energy, $100\,AMeV$, and the
effect appears nicely enhanced, see Fig. \ref{sn132_100_split}. The observed
differences for nucleons at high $p_t$'s already appear 
at early times: since these
momentum regions are populated mostly during the early stages of the reaction
(cfr. \cite{ChenPRC68}), we finally get a permanent signal of a dynamical
feature of the interaction. We note a decreasing behaviour of the N/Z
content of pre-equilibrium emission versus the transverse momentum, 
due to a larger neutron attraction, in the $m^*_n > m^*_p$ case, while
in the opposite case we get a flat, or even increasing, trend.

\begin{figure}[tbp]
\includegraphics[scale=0.5]{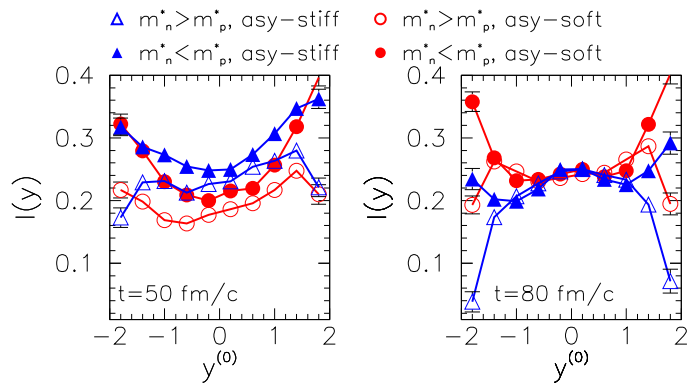}
\caption{\it The asymmetry of fast emitted particles as a function of rapidity,
for the reaction $^{132}Sn +^{124}Sn$ at 100 MeV/A, b=2 fm.} \label{new}
\end{figure}

The N/Z of emitted particles has been studied also as a function of rapidity
(see Fig.\ref{new}). It is possible to see that, at $50\, fm/c$
(end of the pre-equilibrium emission at this beam energy),  differences
between the two prescriptions are present at all rapidities, though
more pronounced for faster particles, as expected. 
We note the opposite behaviour of the two mass splitting sign. 
Going to $80\, fm/c$ (new freeze-out time), the differences at small 
rapidity almost disappear, 
while the different effects already present at $50\, fm/c$ for large rapidities
are kept and are even enhanced. 
The behaviour observed as a function of rapidity is quite similar to 
the trend already seen as a function of the transverse momentum. 
The different mass splitting sign leads to different
results, expecially for particles with larger rapidity or, in
a given rapidity bin, with larger transverse energy.  Using the
$m^*_n<m^*_p$
prescription, the $N/Z$ content of fast emitted particles is enhanced, 
while it is reduced in the opposite case.\\
This is fully consistent with the expectations deduced from the Fig. 
 \ref{lanepseudo} at the end of Sect.II.

Similar results have been obtained in ref. \cite{BaoNPA735} at $400\,AMeV$
in a different $Iso-MD$ model, restricted to the $m_n^*>m_p^*$ choice.

\section{Conclusions}
Our aim has been to find an observable effect of the nucleon effective mass
splitting, which is determined by the momentum dependence of the symmetry
potential. We have performed transport simulations for reactions of 
interest for the
new radioactive beam facilities at intermediate energies.\\
 From our results it appears that the isospin content of pre-equilibrium
emitted nucleons at high transverse momentum, or large rapidity, 
is rather sensitive to this property of the mean field.  Thus it
can be a good observable in order
to learn about fundamental properties of the nuclear interaction, 
such as its momentum dependence in the isovector channel.
In particular, it is found that the  $m_n^*>m_p^*$ splitting leads
to a decreasing behaviour of the $N/Z$ of emitted particles versus
rapidity or transverse energy, while the opposite splitting is associated
with a flat (or even increasing) trend.\\
We finally like to remark that this is an observable of experimental 
interest that can be studied, 
even in absence of information about neutron emissions, by looking
at the correlation between kinematical properties and isotopic content
of light clusters emitted during the early stage of the collision.    





\end{document}